\documentclass[a4paper,11pt]{article}
\pdfoutput=1 

\usepackage{jcappub} 

\usepackage[T1]{fontenc} 

\usepackage{subfig,url}

\title{\boldmath Weak lensing peak statistics in the era of large scale cosmological surveys}

\author[a,1]{J. Fluri,\note{Corresponding author.}}
\author[a]{T. Kacprzak,}
\author[a]{R. Sgier,}
\author[a]{A. Refregier,}
\author[a]{and A. Amara}

\affiliation[a]{Institute for Particle Physics and Astrophysics, Department of Physics, ETH Zurich, \\
Wolfgang-Pauli-Strasse 27, 8093 Zurich, Switzerland}

\emailAdd{janis.fluri@phys.ethz.ch}

\abstract{
Weak lensing peak counts are a powerful statistical tool for constraining cosmological parameters.
So far, this method has been applied only to surveys with relatively small areas, up to several hundred square degrees.
As future surveys will provide weak lensing datasets with size of thousands of square degrees, the demand on the theoretical prediction of the peak statistics will become heightened.
In particular, large simulations of increased cosmological volume are required.
In this work, we investigate the possibility of using simulations generated with the fast Comoving-Lagrangian acceleration (COLA) method, coupled to the convergence map generator \texttt{Ufalcon}, for predicting the peak counts.
We examine the systematics introduced by the COLA method by comparing it with a full TreePM code.
We find that for a 2000 deg$^2$ survey, the systematic error is much smaller than the statistical error.
This suggests that the COLA method is able to generate promising theoretical predictions for weak lensing peaks.
We also examine the constraining power of various configurations of data vectors, exploring the influence of splitting the sample into tomographic bins and combining different smoothing scales.
We find the combination of smoothing scales to have the most constraining power, improving the constraints on the $S_8$ amplitude parameter by at least 40\% compared to a single smoothing scale, with tomography brining only limited increase in measurement precision.

}

\begin{document}
\maketitle
\flushbottom

\section{Introduction}

Gravitational lensing is caused by the deflection of light by the matter between the light source and the observer (see \cite{Bartelmann2001} for a review).
In weak gravitational lensing (WL), the strength of the deflection is small and proportional to the projected mass along the line-of-sight.
This effect can be observed by measuring small, spatially coherent perturbations in the shapes of background galaxies.
As WL probes the projected mass in an unbiased way, it can be used as a powerful cosmological probe.
The potential of this method is shown by weak lensing surveys such as the Canada-France-Hawaii Telescope Lensing Survey (CFHTLenS\footnote{\url{cfhtlens.org}}) \cite{Heymans2013}, the Kilo-Degree Survey (KiDS\footnote{\url{kids.strw.leidenuniv.nl}}) \cite{Hildebrandt2017} or the Dark Energy Survey (DES\footnote{\url{darkenergysurvey.org}}) \cite{Troxel2016}.

In recent years, weak lensing peak statistics have become a increasingly efficient way to analyze the data observed by these surveys (e.g. \cite{Shan2017a, Kacprzak2016a, Liu2014}).
Weak lensing peaks are produced by over-dense regions in the projected mass map and correspond to either individual massive halos or the projection of less massive objects along the line-of-sight \cite{Hamana2004, Liu2016a}. Due to this correspondence to over-dense regions they are an excellent candidate to extract non-Gaussian information from weak lensing data. A tomographic analysis can be done \cite{Martinet2015} by using the redshift information of the background galaxies to create more than one map. This can potentially lead to tighter constraints, provided that the count of background galaxies is high enough to suppress the noise in each redshift bin.
The approaches to create the weak lensing map, on which the peaks are counted, include using aperture mass filter \cite{Schneider1996, Kacprzak2016a}, or the reconstructed convergence maps \cite{Liu2015b}.

Different approaches to analytically predict weak lensing peak counts have been developed over the recent years. For example, high signal-to-noise ratio peaks caused by single massive halos have been studied and used to generate cosmological constraints \cite{Shan2017a}. Other approaches for analytic predictions, such as extreme value statistics, have also been studied \cite{Reischke2016}. Such predictions, while having the advantage of being analytic, can become relatively complex. Another popular way to analyze weak lensing data is to compare the measured peak counts to weak lensing maps generated by N-body simulations \cite{Kacprzak2016a, Dietrich2010, Martinet2017}. Large N-body simulation are, however, computationally intensive and therefore also very time consuming. To reduce the large amount of computing time, emulators like CAMELUS were developed \cite{Lin2015a, Lin2015b, Peel2016a}. CAMELUS is able to generate mock simulations in a fast and efficient way by sampling halos from a mass function and assigning each halo a density profile. While reproducing the peak count well, it lacks precision in some signal to noise regions compared to N-body simulations \cite{Peel2016a}.
 
Further, to use N-body simulations to predict the convergence peak count one has to generate convergence maps out of the simulated matter distribution. Full ray-tracing through the simulations delivers very accurate results, but requires large computational resources. A common way to produce fast convergence maps is to use the first order Born-approximation. However, this approximation breaks down for very small structures and high resolution maps \cite{Petri2017}.  To tackle this issues a novel approach has been developed that generates convergences maps from halos and the linear matter power spectrum \cite{Giocoli2017} . 
 
In our approach, we use the fast, publicly available N-body simulation code L-PICOLA \cite{Howlett2015} to predict the peak counts for various cosmologies. L-PICOLA has the advantage of producing accurate approximations of full N-body simulations in a fraction of the time \cite{Howlett2015, Sgier2018}. The \texttt{Ufalcon} package, a fast cosmological map making pipeline \cite{Sgier2018}, was then used to generate the convergence maps from the L-PICOLA outputs. We show that these simulations are sufficiently accurate to be used as mock catalogs for weak lensing peak statistics. This was done by comparing L-PICOLA to the full TreePM code Gadget-2 \cite{Springel2005}. We show the potential of this approach in large scale surveys to constrain the total matter density $\Omega_m$ and the fluctuation amplitude $\sigma_8$ with peak statistics. Further we explore the impact of different Gaussian smoothing kernels and tomographic redshift bins on the constraints.

We also include noise in the weak lensing maps which needs to be properly corrected for in the peak analysis. Specifically, we include shape noise and the measurement noise in the galaxy shapes, which have been extensively studied (e.g. \cite{bard2017}). Other effects include the magnification bias which is caused by preferential selection of magnified source galaxies \cite{Liu2013a}, effects of observational masks \cite{bard2017a, Liu2013}, biases that are caused by inverting cosmic shear maps to convergence maps \cite{Lin2017} and the intrinsic alignment of galaxies around large clusters \cite{Heavens2000}. In our approach we did not include these systematic effects, as we focused on the theoretical aspects of the peak count predictions.

This paper is structured in the following way. In section \ref{s:methodology}, we give an overview of all methods used in this paper. Section \ref{s:nobody} covers our generated N-body simulations. In section \ref{s:gen_convergence}, we explain how we generated our convergence maps. How we convert these convergence maps into cosmic shear maps is explained in section \ref{s:shearconversion}. The procedure of adding noise and the actual peak measurement is covered by sections \ref{s:addnoise} and \ref{s:peak_measurement}. Our likelihood analysis is done in section \ref{s:loglike}. In section \ref{s:coscons}, we present the resulting cosmological constraints of our non-tomographic and tomographic analysis, which is followed by our conclusion in section \ref{s:conlusion}. Appendix \ref{s:settings} gives further insight into our used L-PICOLA settings and in appendix \ref{s:convergence} we explain the convergence map generation in more detail. Finally appendix \ref{s:cherrypicking} explains how we chose our mock observation for our cosmological constraints.

\section{Method}
\label{s:methodology}

We started by generating N-body simulations of 70 different cosmologies in the $\Omega_m$-$\sigma_8$ plane and used them to generate full sky convergence maps, using \texttt{Ufalcon} \cite{Sgier2018}. These convergence maps were then transformed into cosmic shear maps using spherical harmonics decomposition \cite{Wallis2017}. Out of each full sky shear map we cut out multiple 1735 square degrees mock surveys. Noise was then added and the masked, noisy maps were transformed back to convergence maps. To reduce the noise we smoothed the noisy convergence maps with Gaussian filters of various scales. We used these mock surveys to build a likelihood and used it to constrain these two cosmological parameters, using one of the simulations as our mock observation.

\subsection{N-Body Simulations}
\label{s:nobody}

The N-body simulations were done using the fast L-PICOLA code \cite{Howlett2015}. This is a distributed memory, planar-parallel code using the comoving Lagrangian acceleration method (COLA method) making it several orders of magnitude faster than an equivalent full N-body simulation. L-PICOLA solves the necessary differential equations with a particle mesh algorithm. These approximation lead to a lower accuracy in small scale clustering. To measure the systematic limits of this approximation, we further used one full N-body simulation for our fiducial cosmology generated by the publicly available Gadget-2 code \cite{Springel2005}. The used simulation was the same as in \cite{Sgier2018}.

One great advantage of the L-PICOLA code is its ability to generate lightcone simulations.
This leads to a substantial reduction in the simulation output, as it does not require storing the full particle distribution for all time steps.
Using the lighcone mode therefore makes it possible to generate lightcone simulations with a good redshift resolution with minimal memory usage. However since the Gadget-2 code does not have this feature, all simulations were also run in snapshot mode to get a fair comparison. In this work we were mainly interested in the effects of the COLA approximation on the cosmological constraints. How we measured this effects and how these systematics affected the constraints is explained in sections \ref{s:loglike} and \ref{s:coscons}.

\begin{figure}[]
\centering
\includegraphics[width=0.75\textwidth]{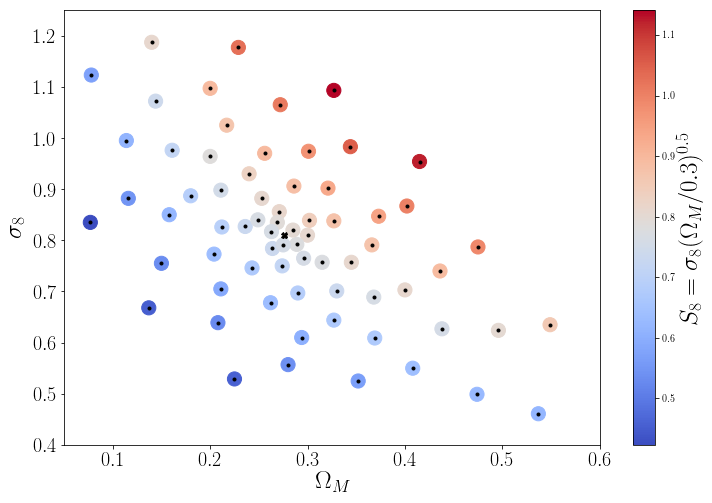}
\caption{Grid of simulated cosmologies. The color is given by the degeneracy parameter $S_8$. The cross in the middle represents our fiducial cosmology. \label{f:simugrid}}
\end{figure}

Assuming a flat $\Lambda CDM$ universe the parameters of our fiducial cosmology were set to $\Omega_m = 0.276$, $\sigma_8 = 0.811$, $h = 0.7$, $n_s = 0.961$ and $\Omega_\mathrm{baryon} = 0.045$. We generated a total of ten full sky simulations of our fiducial model. Further 69 simulations were done changing only $\Omega_m$ and $\sigma_8$ (and $\Omega_\Lambda$ such that the universe remains flat). The whole simulation grid is shown in figure \ref{f:simugrid}. The simulation grid is non-uniform and denser around the fiducial cosmology to properly sample the region of degeneracy. All points of the simulation grid lay on ellipses $\mathcal{E}(t) = (a\cos(t), b\sin(t))$ centered around our fiducial cosmologies and for each ellipse we chose equally spaced values of $t \in [0, 2\pi)$. The density of the ellipses increases towards the fiducial cosmology and we only considered points inside our prior range of $\Omega_m \in [0, 0.7]$ and $\sigma_8 \in [0.4, 1.4]$.

The Gadget-2 simulations of our fiducial cosmology, from \cite{Sgier2018}, were done using $1024^3$ particles and started at redshift $z=50$. Snapshots were outputted from redshift $z=1.5$ to redshift $z=0$ with a step size of $\Delta z = 0.01$. To generate a lightcone with redshift depth of $z = 1.5$, the box size needs to have a length of approximately 9 Gpc. However, to get a better particle resolution in the lower redshifts, two simulations were done: one with a box size of 6 Gpc and one with a box size of 9 Gpc.

As suggested in \cite{Howlett2015}, we started our L-PICOLA simulations at redshift $z=9$ and snapshots were outputted in the same redshift intervals as for the Gadget-2 simulations. However, to save computational time we used smaller boxes and used them multiple times as periodic boxes. The box sizes where therefore only one fourth the box sizes simulated with the Gadget-2 code. To keep the number of particles per volume the same, the L-PICOLA simulations used $256^3$ particles and a mesh with $1024^3$ points. Each 6 Gpc and 9 Gpc box was generated out of 64 periodic boxes. As in \cite{Dietrich2010},  we applied random shifts, rotations and parity flips on each of these 64 boxes, to reduce the resulting spatial correlation. Further details about how we chose our L-PICOLA settings can be found in appendix \ref{s:settings}.

\subsection{Generating Convergence Maps}
\label{s:gen_convergence}

The generation of the full sky convergence maps was done using \texttt{Ufalcon} as in \cite{Sgier2018}, which follows the appendix of \cite{Teyssier2009} and uses the Hierarchical Equal Area iso-Latitude Pixelization tool\footnote{http://healpix.sourceforge.net} ($\mathtt{HEALPix}$ \cite{Healpix}). A quick overview of this method is given in appendix \ref{s:convergence}. Out of each snapshot, we cut out a shell of thickness $\Delta z = 0.01$ to generate a past-lightcone. As in \cite{Sgier2018}, the approach with nested boxes allowed us to generate a lightcone from $z=0.1$ to $z=1.5$ using the 6 Gpc boxes for redshifts $z = 0.1$ to $z = 0.79$ and the 9 Gpc boxes for redshift $z = 0.8$ to $z = 1.49$.

This past-lightcone was then used to generate full sky convergence maps with a realistic redshift distributions given by \cite{Smail1994}
\begin{equation}
n(z) \propto z^2\exp\left(-\frac{z}{0.2}\right), \label{eq:nz1} \\
\end{equation}
which is shown in figure \ref{f:tomo} with the solid black line. All generated maps had an $\mathtt{HEALPix}$ nside of 1024.

\subsection{Generating Shear Maps}
\label{s:shearconversion}

To convert the generated full sky convergence maps into cosmic shear maps we used the spherical harmonics decomposition. The convergence can be decomposed in the following way
\begin{equation}
\kappa(\theta,\phi) = \sum_{l=0}^\infty\sum_{m = -l}^l\hat{\kappa}_{lm}Y_{lm}(\theta,\phi),
\end{equation}
where the $\hat{\kappa}_{lm}$ are the spherical harmonic coefficients and the $Y_{lm}(\theta,\phi)$ are the usual spin 0 spherical harmonics. The cosmic shear field on the other hand is a spin 2 field, which means it decomposes as
\begin{equation}
\gamma(\theta,\phi) = \sum_{l=0}^\infty\sum_{m = -l}^l{}_2\hat{\gamma}_{lm}{}_{2}Y_{lm}(\theta,\phi),
\end{equation}
where the $_{2}Y_{lm}(\theta,\phi)$ are the spin 2 spherical harmonics. As shown in \cite{Wallis2017}, one can connect the coefficients of convergence and shear over the gravitational lensing potential to get the following relation
\begin{equation}
{}_2\hat{\gamma}_{lm} = \frac{-1}{l(l+1)}\sqrt{\frac{(l+2)!}{(l-2)!}}\hat{\kappa}_{lm}.
\label{eq:kappashear}
\end{equation}
 A similar procedure was also done in \cite{Chang2017}. Using this equation and the routines implemented in $\mathtt{HEALPix}$ we generated full sky cosmic shear maps out of the convergence maps. Usually one decomposes the convergence and cosmic shear fields into an E-mode with even parity and a B-mode with odd parity. The B-modes of both fields should theoretically vanish. However if one considers not the full sky, but masked areas and noisy maps it is possible to obtain non-vanishing B-modes. In this work, we only considered E-modes and neglected the B-modes. We found that by using equation \ref{eq:kappashear} combined with a Gaussian smoothing kernel (see section \ref{s:addnoise}) it is possible to convert maps with almost no loss of information.

\subsection{Shape and Measurement Noise}
\label{s:addnoise}

In order to add noise to the generated shear maps we produced a galaxy catalog of around 19 million galaxies according to the redshift distribution given by equation \ref{eq:nz1}. These galaxies were uniformly distributed over a 1735 square degree area. To connect the catalog to a specific simulation we assigned each galaxy $i$ a true shear value $\gamma^\mathrm{True}_i$ according to the pixel $p(i)$ it would fall into on the generated shear map $\gamma^\mathrm{map}_{p(i)}$
\begin{equation}
\gamma^\mathrm{True}_i = \gamma^\mathrm{map}_{p(i)}.
\end{equation}
For each galaxy a noise term $e^\mathrm{noise}_i$ was generated by drawing a random number $\vert e \vert_i$ according to the noise distribution given by equation \ref{eq:noise_dist} and randomly rotating it
\begin{equation}
e^\mathrm{noise}_i = \vert e \vert_i\exp(i*\varphi),
\end{equation}
where $\varphi \in [0,2\pi]$ was drawn from a uniform distribution. The noise distribution is parametrized by the equation
\begin{equation}
p(\vert e \vert) \propto \left(\vert e \vert + 0.01 \right)^{-5}\left(1 - \exp\left(-70|e|^{4.93}\right)\right),
\label{eq:noise_dist}
\end{equation}
and represents a shape and measurement noise distribution. This was chosen to resemble $p(\vert e \vert)$ from \texttt{UFig} simulations, similar to the ones in \cite{UFig}. The estimated shear value for a given galaxy $i$ was then obtained by simply adding the noise term to the true shear value
\begin{equation}
\gamma^\mathrm{est}_i = \gamma^\mathrm{True}_i + e^\mathrm{noise}_i.
\end{equation}
To convert this catalog back to a map the estimated shear values of the galaxies were averaged over the pixels
\begin{equation}
\gamma^\mathrm{map, noisy}_j = \frac{1}{N_j}\sum_{p(i) = j}\gamma^\mathrm{est}_i,
\end{equation}
where $\gamma^\mathrm{map, noisy}_j$ is the shear value of the $j$th pixel and $N_j$ is the number of galaxies that fall into this pixel. Since the used mask covers only a small fraction of the sky, the catalog was rotated to generate a total of 13 patches with the same mask out of one full sky simulation.
\begin{figure}[]
\centering
\includegraphics[width=1.0\textwidth]{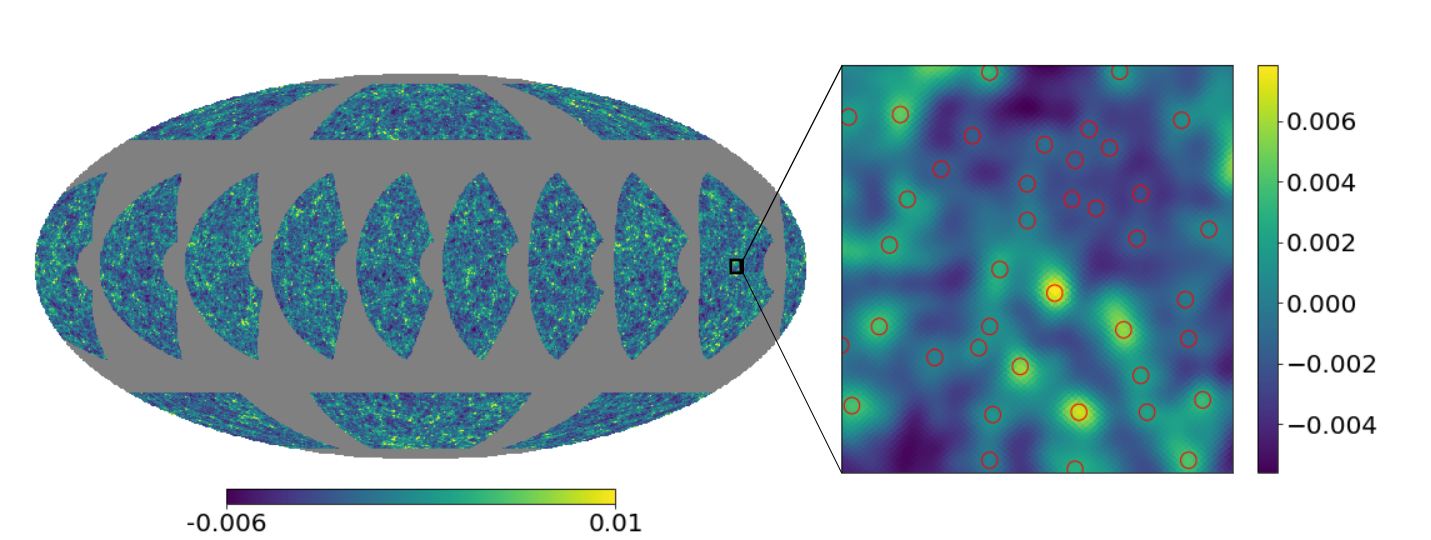}
\caption{Left: A noise free convergence map smoothed with a Gaussian smoothing kernel with FWHM = 21.1 arcmin where the masks of the 13 patches that were used to generate noisy shear maps are shown. Each patch was used separately to generate one peak measurement. The patches were obtained by rotating the position of the catalog galaxies and projecting them on the sphere. Right: The black $5 \times 5$ square degree patch on the left side enlarged. All peaks are highlighted by red circles.}
\label{f:patches}
\end{figure}
And for each of these patches we produced 20 independent noisy realizations for the non-tomographic run and 30 for the tomographic run (see section \ref{s:coscons}). We also produced noise only shear maps for each noisy realization to count the random peaks. The  patches are shown in figure \ref{f:patches}. Afterwards the maps were converted back to convergence maps using the formalism explained in section \ref{s:shearconversion}. However, since the 13 generated patches cover only a fraction of the full sky, the spherical harmonics decomposition leads to some boundary effects and thus to larger errors at the edge of the patches. Therefore, we removed a small number of pixels around the edge of the patches to counter these boundary effects. To decide which pixels should be removed we generated one full sky map for each of the 13 patches where we set the values of all pixels inside the patch to 1 and all other to 0. Afterwards we performed a spherical harmonics decomposition on each of the maps and applied a large Gaussian smoothing kernel with FWHM = 21.1 arcmin and transformed them back to full sky maps. We decided to remove all pixels inside the patch where the values of the pixels deviated more than 5\% from their original value before we performed the transformation. 
After removing the boundary pixels from the converted noisy maps, the maps were smoothed with a Gaussian kernel. We applied 12 different scales FWHM = \{31.6, 29, 26.4, 23.7, 21.1, 18.5, 15.8, 13.2, 10.5, 7.9, 5.3, 2.6\} arcmin on each noisy realization. The smoothing scales were chosen in such a way that increasing their number did not lead to a significant improvement of the cosmological constraints.

\subsection{Peak Measurement}
\label{s:peak_measurement}

\begin{figure}[]
\centering
\includegraphics[height=53mm]{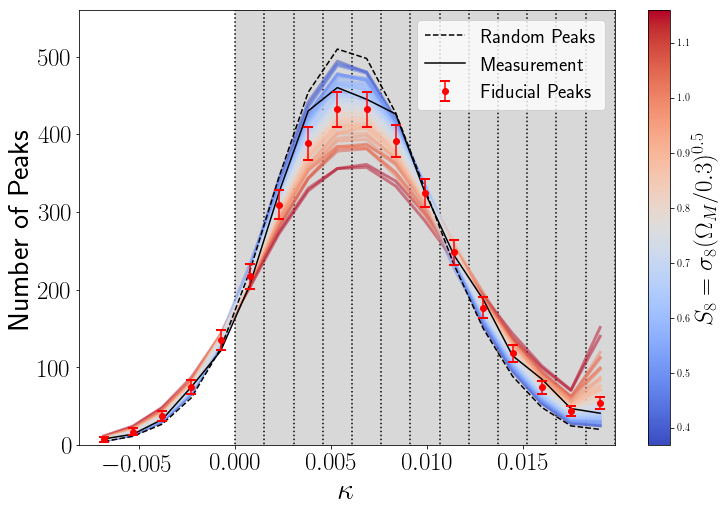}
\includegraphics[height=53mm]{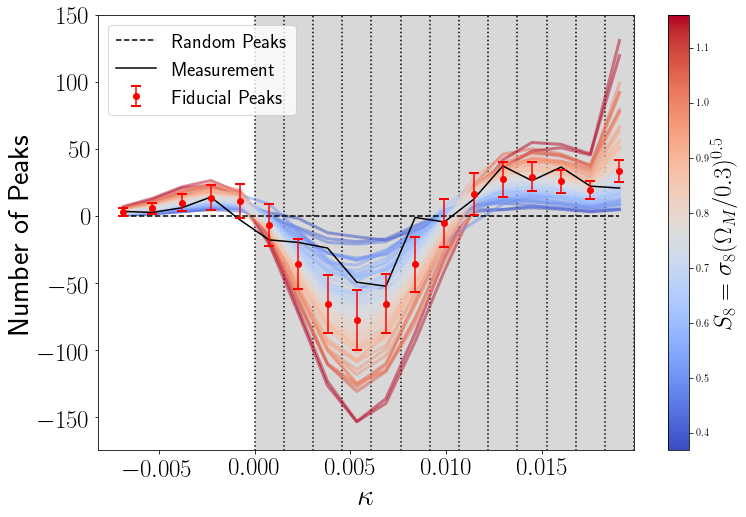}
\caption{Average peak counts from all simulations for a smoothing scale of FWHM = 21.1 arcmin (non tomographic). The color denotes the degeneracy parameter $S_8$. The black solid line denotes our mock observation. The random peaks show the average number of peak counts obtained from convergence maps that contained only shape noise. In the right panel they were subtracted from all average peak counts. The fiducial peaks are the average number of peak counts from the 2600 noisy realizations from our fiducial cosmology and the error bars show the standard deviation. The grey highlighted regions show the bins that were used for the likelihood analysis.}
\label{f:peaks}
\end{figure}

As in \cite{Liu2015b}, we counted the peaks on the actual convergence map. A peak is defined as a pixel that has a higher convergence value than all its eight neighboring pixels. This is shown on the right side of figure \ref{f:patches}. These peaks were then binned according to their value and the number of peaks in each bin was counted. We used a variation of the linear binning scheme: we used 12 linearly spaced bins and added a 13th bin without an upper boundary. Therefore all convergence peaks above the second to last bin fell automatically into the last bin and were used for the likelihood analysis. An example of such peak functions is shown in figure \ref{f:peaks}.
We only considered peaks with a positive convergence value as done in \cite{Kacprzak2016a} and neglected peaks with a negative convergence value. As seen in figure \ref{f:peaks}, this peak function range only carries little information.

\subsection{Likelihood Analysis}
\label{s:loglike}
The likelihood analysis was done in the same way as in \cite{Kacprzak2016a}. We assumed that the measured vector $\mathbf{\hat{d}}$, containing the number of peaks in each bin, fluctuates due to noise and cosmic variance around its mean value $\mathbf{d}$ with a multivariant Gaussian distribution
\begin{equation}
\mathbf{\hat{d}}\sim\mathcal{N}\left(\mathbf{d},\Sigma\right),
\end{equation}
where $\Sigma$ is the covariance matrix. We estimated the covariance matrix using our fiducial simulations
\begin{equation}
\hat{\Sigma} = \frac{1}{N_s-1}\sum_{i=1}^{N_s}\left(\hat{\mathbf{d}}_i-\bar{\mathbf{d}}\right)\left(\hat{\mathbf{d}}_i-\bar{\mathbf{d}}\right)^T,
\end{equation}
where $N_s=2600$ ($N_s = 3900$ for the tomographic run) is the number of noisy realizations, $\hat{\mathbf{d}}_i$ is the vector of counted peaks of the $i$th noisy realization and $\bar{\mathbf{d}}$ is the vector containing the average peak counts of all noisy realizations.
The correlation matrix is defined by
\begin{equation}
\hat{\Sigma}^\mathrm{corr}_{ij} = \frac{\hat{\Sigma}_{ij}}{\sqrt{\hat{\Sigma}_{ii}\hat{\Sigma}_{jj}}} \in [-1,1].
\end{equation}
which was used for visualization, since it has a well defined dynamic range.
As mock observation $\mathbf{\hat{d}}_\mathrm{meas.}$ we always used one noisy realization of the fiducial cosmology. How we picked this mock observation is explained in appendix \ref{s:cherrypicking}. We denote $\mathbf{d}(\mathbf{\pi})$ as the vector of the average number of peak counts from the simulated cosmology $\mathbf{\pi} = \{\Omega_m, \sigma_8\}$. Since we did not consider any systematics, like intrinsic alignment or boost factors as in \cite{Kacprzak2016a}, our parameter space was chosen to be two dimensional. We always chose our convergence bins in a way that each bin contained at least 25 peak counts. Assuming that each bin has at least this number of peak counts one can build a Gaussian likelihood analysis. Using a Bayesian approach with flat priors $\Omega_m \in [0, 0.7]$ and $\sigma_8 \in [0.4, 1.4]$ the probability of measuring $\hat{\mathbf{d}}_\mathrm{meas.}$ if the true parameters are $\mathbf{\pi}$, is then proportional to
\begin{equation}
p(\hat{\mathbf{d}}|\mathbf{\pi})\propto \exp\left(-\frac{1}{2}\frac{N_s-N_d-2}{N_s-1}\left(\hat{\mathbf{d}}_\mathrm{meas.}-\mathbf{d}(\mathbf{\pi})\right)^T\hat{\Sigma}^{-1}\left(\hat{\mathbf{d}}_\mathrm{meas.}-\mathbf{d}(\mathbf{\pi})\right)\right),
\end{equation}
where $N_d$ is the length of the vector $\hat{\mathbf{d}}_\mathrm{meas.}$ and $\hat{\Sigma}^{-1}$ is the inverse of our covariance estimate. The prefactor was calculated according to \cite{Hartlap2007} and ensures that we obtain an unbiased estimate of the inverse covariance matrix

To get the likelihood of points outside of our simulation grid we created different interpolation schemes. We found that a stable and efficient way to interpolate the single bins was to use smooth bivariant splines\footnote{\texttt{SmoothBivariateSpline} from the \texttt{scipy.interpolate} package}. These were used for likelihood analysis following below and were consistent with the scheme used in \cite{Kacprzak2016a}.

\subsubsection{Influence of the COLA Approximation}

\begin{figure}[]
\centering
\includegraphics[width=0.48\textwidth]{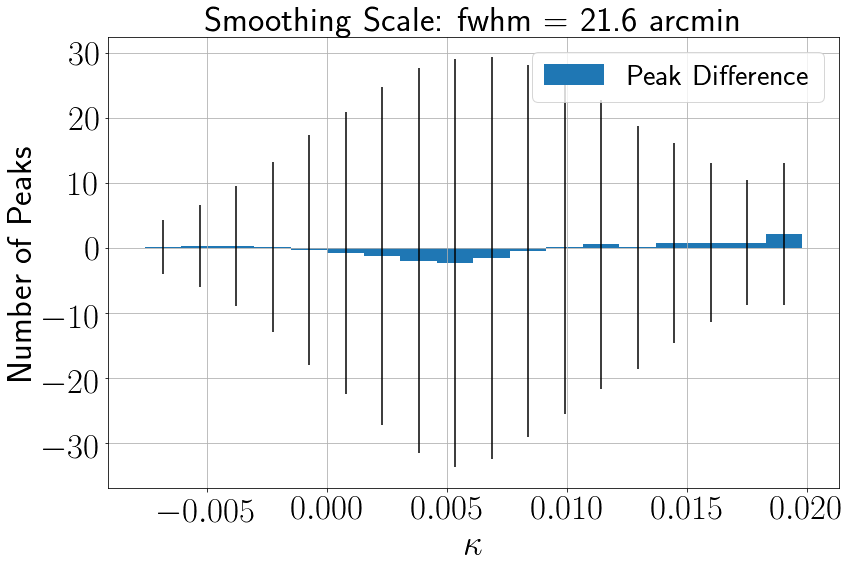}
\includegraphics[width=0.48\textwidth]{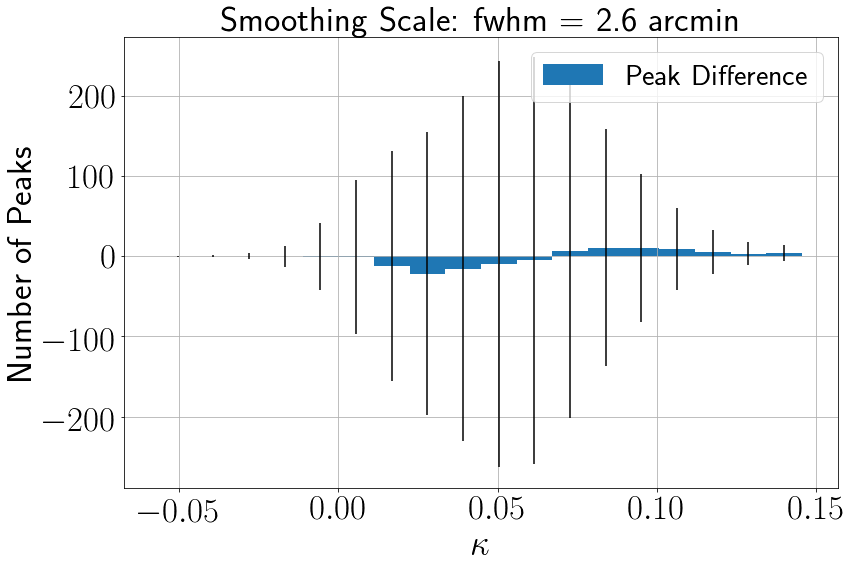}
\caption{Average peak count difference of 2600 L-PICOLA and Gadget-2 noisy realizations (non tomographic). The error bars correspond to the added standard deviation of the peak counts from the Gadget-2 and L-PICOLA noisy realizations. A smoothing scale of 21.1 arcmin with the same bins as in figure \ref{f:peaks} is shown on the left side. On the right side a our smallest smoothing scale of 2.6 arcmin was used. \label{f:diff}}
\end{figure}

To measure the influence of the COLA approximation we generated 200 noisy realizations out of each patch from our Gadget-2 simulation. We calculated the mean peak count difference of the Gadget-2 and L-PICOLA simulations
\begin{equation}
\Delta \bar{\mathbf{d}} = \bar{\mathbf{d}}_\mathrm{Gadget-2} - \bar{\mathbf{d}}_\mathrm{L-PICOLA}, \label{eq:sys}
\end{equation}
where $\bar{\mathbf{d}}_\mathrm{Gadget-2}$ and $\bar{\mathbf{d}}_\mathrm{L-PICOLA}$ are the vectors containing the average number of peak counts of all noisy realizations from Gadget-2 and L-PICOLA from our fiducial cosmology. An example of such a difference is shown in figure \ref{f:diff}.
The smoothing scale used for this plot was 21.1 arcmin (left) and 2.6 arcmin (right). Other smoothing scales showed a similar level of difference. Even though the difference is small compared to the total number of peaks in each bin, a small bias, which also depends on the used smoothing scale, is visible. To propagate this bias caused by the COLA approximation to the cosmological constraints, the difference from equation \ref{eq:sys} was added to mock observation and a second likelihood analysis was performed as a tolerance analysis. One should note that the calculated bias may be depending on the choice of the fiducial cosmology. However, we do not think that this bias will drastically change in the neighborhood of our fiducial cosmology.

\section{Cosmological Constraints}
\label{s:coscons}

We compared the constraints from the different settings in two ways. The first approach was to compare their constraining power on the degeneracy parameter $S_8 = \sigma_8\left(\Omega_m/0.3\right)^{0.5}$. Our second approach was to compare the figure-of-merit (FoM). For this we used the same definition of the FoM as \cite{Troxel2016}, which is defined for two parameters $p_1$ and $p_2$ as
\begin{equation}
\mathrm{FoM}_{p_1-p_2} = \frac{1}{\sqrt{\det\left(\mathrm{Cov}(p_1,p_2)\right)}}.
\end{equation}

\subsection{Non-Tomographic Constraints}

\begin{figure}[]
\centering
\includegraphics[height=67mm]{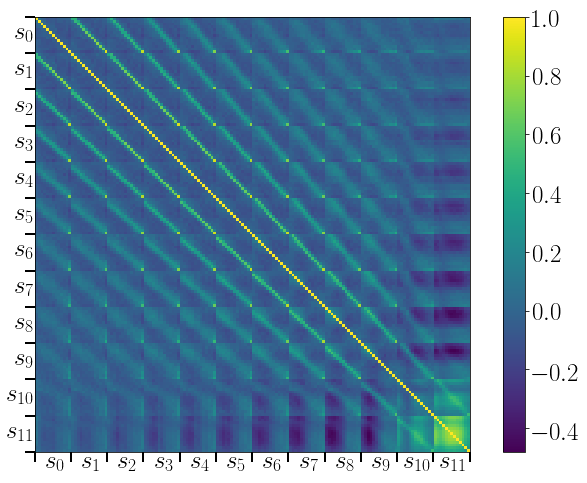}
\includegraphics[height=67mm]{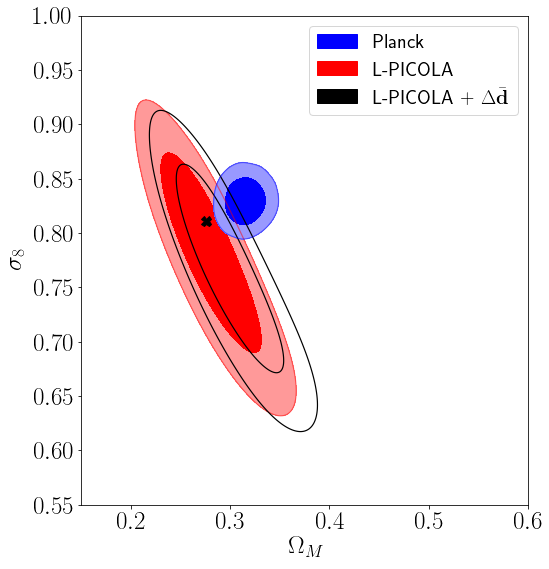}
\caption{Left: Correlation matrix obtained by combining all 12 smoothing scales, generated from the peak counts of 2600 noisy realizations. It is ordered from the largest smoothing scale $s_0$ (upper left) to smallest smoothing scale $s_{11}$(lower right). Each smoothing scale contains 13 bins of convergence peaks. Right: Cosmological constraints obtained by combining all 12 smoothing scales. The red contours include the 0.68 and 0.95  confidence region from our mock observation. For the black contours we added the systematics described by equation \ref{eq:sys}. The black cross shows our fiducial cosmology. For comparision the blue contours show the results obtained from Planck. Note that the fiducial cosmology in this work is not centered at the Planck result. Therefore the position of the contours should not be compared, only their relative sizes.  \label{f:nz1singlesmooth}}
\end{figure}

The smoothing scale used has significant effect on the constraints. A large smoothing scale will smooth out most of the noise and the effects of the COLA approximation. However, it will also reduce the cosmological signal, which will lead to larger errors. If the smoothing scale is chosen too low, the effects of the COLA approximation will start to become significant. This can be seen by comparing the two panels of figure \ref{f:diff}.
Since different smoothing scales correspond to objects of different sizes, it is possible to improve the constraints by combining different smoothing scales. The combination of smoothing scales was done by combining the data vectors of the single smoothing scales to a single vector. In figure \ref{f:nz1singlesmooth} the correlation matrix and the constraints obtained by combining all our 12 different smoothing scales are shown by the red contours. An overview of different settings and the resulting $S_8$ constraints and figures-of-merit is given in table \ref{t:nontomo}. The choice of our mock observation is explained in appendix \ref{s:cherrypicking}. The black contours correspond to the same measurement, except that the systematics coming from the COLA approximation were added. The COLA approximation adds a bias of approximately $\frac{1}{3} \sigma$.
\begin{table}[]
\centering
\begin{tabular}{|c|c|c|c|}
\hline
FWHM & $S_8$ L-PICOLA & $S_8$ with offset & $\mathrm{FoM}_{\sigma_8-\Omega_m}$ \\
\hline
& & & \\[-0.8em]
31.6 arcmin & $0.74^{+0.04}_{-0.05}$ & $0.75^{+0.05}_{-0.05}$ & 155.58 \\[0.2em]
\hline
& & & \\[-0.8em]
15.8 arcmin & $0.74^{+0.05}_{-0.07}$ & $0.74^{+0.05}_{-0.06}$ & 303.61 \\[0.2em]
\hline
& & & \\[-0.8em]
2.6 arcmin & $0.75^{+0.07}_{-0.12}$ & $0.79^{+0.06}_{-0.11}$ & 275.97 \\[0.2em]
\hline
& & & \\[-0.8em]
combination of 12 & $0.76^{+0.03}_{-0.03}$ & $0.77^{+0.03}_{-0.04}$ & 947.81 \\[0.2em]
\hline
\hline
\multicolumn{4}{|c|}{} \\[-0.8em]
\multicolumn{4}{|c|}{Planck $\mathrm{FoM}_{\sigma_8-\Omega_m}$: 5265.86} \\[0.2em]
\hline
\end{tabular}
\caption{Table of the non tomographic $S_8$ constraints and figures-of-merit. The $S_8$ value corresponds to the peak of its pdf and the uncertainties to the 68\% confidence interval. The figure-of-merit was calculated without the systematics described by equation \ref{eq:sys}.\label{t:nontomo}}
\end{table}
As a point of reference, we compared our results to constraints obtained from Planck \cite{PC2015} (available on the Planck legacy archive\footnote{https://pla.esac.esa.int/pla/{\#}home}). They were obtained from CMB temperature and polarization measurements (using temperature ($\ell = 30 -2508$) and temperature+polarization ($\ell = 2 - 29$)). One should note that the position of the confidence areas are not comparable since the contours of Planck were generated using real data and our contours were generated using our mock observation.

\subsection{Tomographic Constraints}

\begin{figure}[]
\centering
\includegraphics[width=0.75\textwidth]{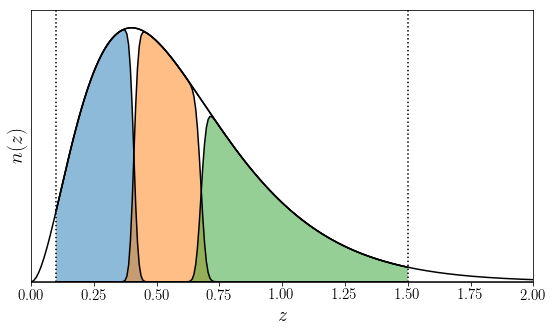}
\caption{The three redshift distributions, given by the blue, red and green area, used to generate the convergence maps for our tomographic run. The redshift distribution $n(z)$, given by equation \ref{eq:nz1} and denoted by the solid black line, is wrapped around them.}
\label{f:tomo}
\end{figure}

\begin{figure}[]
\centering
\includegraphics[height=67mm]{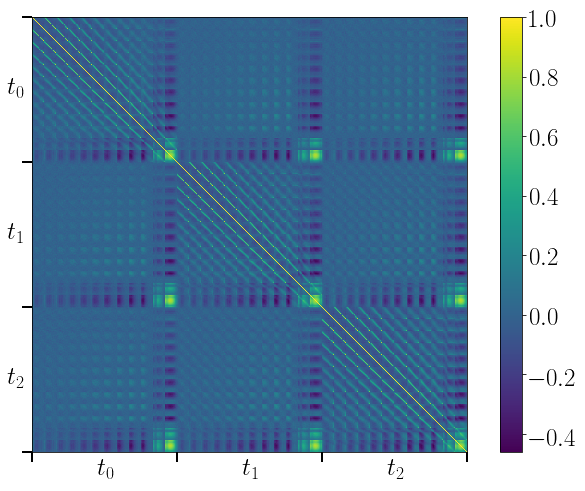}
\includegraphics[height=67mm]{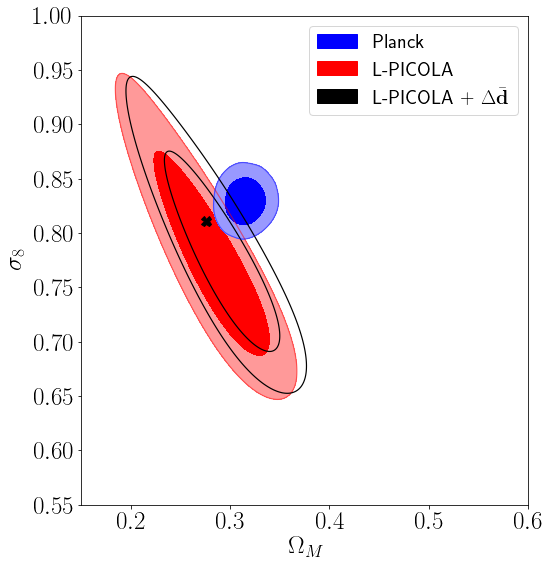}
\caption{Left: Correlation matrix obtained by combining all 12 smoothing scales and all three redshift bins, generated from the peak counts of 3900 noisy realizations. It is ordered from the first redshift bin $t_0$ (upper left) to the third redshift bin $t_2$ (lower right). Each tomographic bin is made out of the combination of all 12 smoothing scales and is ordered in the same way as the correlation matrix in figure \ref{f:nz1singlesmooth}. Right: Cosmological constraints obtained by combining all 12 smoothing scales and all three redshift bins. The red contours include the 0.68 and 0.95  confidence region from our mock observation. For the black contours we added the systematics described by equation \ref{eq:sys}. The black cross shows our fiducial cosmology. The blue contours show the results obtained from Planck. Note that the fiducial cosmology in this work is not centered at the Planck result. Therefore the position of the contours should not be compared, only their relative sizes. \label{f:nz1tomo}}
\end{figure}

A tomographic analysis uses the redshift information of the background galaxies to create multiple maps. Each maps is generated by considering only galaxies in a given redshift range. For this purpose, we simulated a measurement error for the true redshift of our galaxy catalog. For each galaxy we sampled a random number $\Delta z_\mathrm{true}$ from a normal distribution $\mathcal{N}(z_t, \delta(1+z_t))$, where $z_t$ is the true redshift of the galaxy and we set $\delta = 0.01$. We then shifted the true redshift $z_t$, for each galaxy, by these random values and assumed it to be the photometric redshift of the galaxy. We then split the catalog into 3 equal partitions according to their photometric redshift. We recovered the true redshift distribution of the photometric redshift bins according to \cite{Amara2007}. The used true redshift distributions of the 3 bins are shown in figure \ref{f:tomo}.
We then generated 3 full sky convergence maps for each simulation according to the redshift distributions of the bins. The noise was then added using only galaxies that would fall into the corresponding redshift bin. We then performed the same likelihood analysis as above. The correlation matrix and the cosmological constraints obtained by combining all our 12 smoothing scales are shown in figure \ref{f:nz1tomo}. We used the same Planck results as above as point of reference. The $S_8$ constraints and the figures-of-merit from different setting are listed in table \ref{t:tomo}.
\begin{table}[]
\centering
\begin{tabular}{|c|c|c|c|}
\hline
FWHM & $S_8$ L-PICOLA & $S_8$ with offset & $\mathrm{FoM}_{\sigma_8-\Omega_m}$ \\
\hline
& & & \\[-0.8em]
31.6 arcmin & $0.75^{+0.05}_{-0.05}$ & $0.77^{+0.04}_{-0.05}$ & 175.34 \\[0.2em]
\hline
& & & \\[-0.8em]
15.8 arcmin & $0.71^{+0.05}_{-0.05}$ & $0.74^{+0.05}_{-0.05}$ & 287.39 \\[0.2em]
\hline
& & & \\[-0.8em]
2.6 arcmin & $0.66^{+0.08}_{-0.05}$ & $0.67^{+0.09}_{-0.04}$ & 241.97 \\[0.2em]
\hline
& & & \\[-0.8em]
combination of 12 & $0.75^{+0.03}_{-0.03}$ & $0.77^{+0.03}_{-0.03}$ & 890.84 \\[0.2em]
\hline
\hline
\multicolumn{4}{|c|}{} \\[-0.8em]
\multicolumn{4}{|c|}{Planck $\mathrm{FoM}_{\sigma_8-\Omega_m}$: 5265.86} \\[0.2em]
\hline
\end{tabular}
\caption{Table of the tomographic $S_8$ constraints and figures-of-merit. The $S_8$ value corresponds to the peak of its pdf and the uncertainties to the 68\% confidence interval. The figure-of-merit was calculated without the systematics described by equation \ref{eq:sys}. \label{t:tomo}}
\end{table}
The average of the tomographic constraints is similar, but slightly worse, than for the non-tomographic case. For this case, tomography did not improve the constraints. However, increasing the galaxy count to reduce the noise in each redshift bin or considering spatial correlations between the redshift bins, could potentially improve these constraints. It is therefore possible to find tighter constraints by using tomography which, for example, was the case in \cite{Petri2016}.

\section{Conclusion and Outlook}
\label{s:conlusion}

In this work, we explored the possibility of using fast L-PICOLA method, coupled to the convergence map generator \texttt{Ufalcon} \cite{Sgier2018}, as a theory prediction tool for weak lensing peak statistics, as well as the impact of the choice of the data vector on the cosmological constraints.
To test the validity of the approximation introduced by L-PICOLA, we simulated cosmologies using L-PICOLA and compared it to the full TreePM code Gadget-2.
Using these simulations, we were able to generate cosmological constraints.
By doing this we have successfully shown that it is possible to use the fast L-PICOLA code to constrain cosmological parameters via weak lensing peak statistics.
The systematics caused by the COLA approximation depend mainly on the used smoothing scale and on the signal to noise level.
These systematics are, however, small compared to the statistical uncertainties, on the level of $\frac{1}{3}\sigma$ on $S_8$ parameter.

Further, we were able to significantly improve the cosmological constraints by combining different smoothing scales.
The uncertainty on the $S_8$ parameter was at least $40\%$ smaller than using a single smoothing scale.
We performed a tomographic likelihood analysis and unexpectedly found that this did not improve our constraints significantly.
This result can be explained by the low galaxy count in our redshift bins. We suspect that the information added in the tomographic analysis is balanced by the increased noise due to the lower number of galaxies in the different redshift bins. One should also note that we did not consider any systematics.
In a real application, systematics like intrinsic alignment could potentially be broken by performing a tomographic analysis.

One obvious extension of this work would be to use this tool on real data.
To do this it would be necessary to model various systematics like intrinsic alignment or baryonic effects.
Another improvement would be to run \texttt{Ufalcon} with L-PICOLA in lightcone mode, as it is described in \cite{Sgier2018}. This would reduce to memory usage and could potentially improve the redshift resolution of the lightcone.

Another way to potentially improve the constraints would be to combine weak lensing peak statistics with other probes.
The two point correlation function for example does not extract the same information from the weak lensing maps. Therefore combining this two statistics one could possible achieve better results.

\acknowledgments

This work was support in part by grant number 200021\_169130 from the Swiss National Science Foundation.
We acknowledge the support of IT services of Euler cluster at ETH Zurich.
TK would like to thank Joerg Dietrich and Zoltan Haiman for helpful discussions.
We thank Jose Manuel Zorrilla Matilla for valuable comments.

\appendix

\section{L-PICOLA Settings}
\label{s:settings}

\begin{figure}[]
\centering
\includegraphics[width=0.75\textwidth]{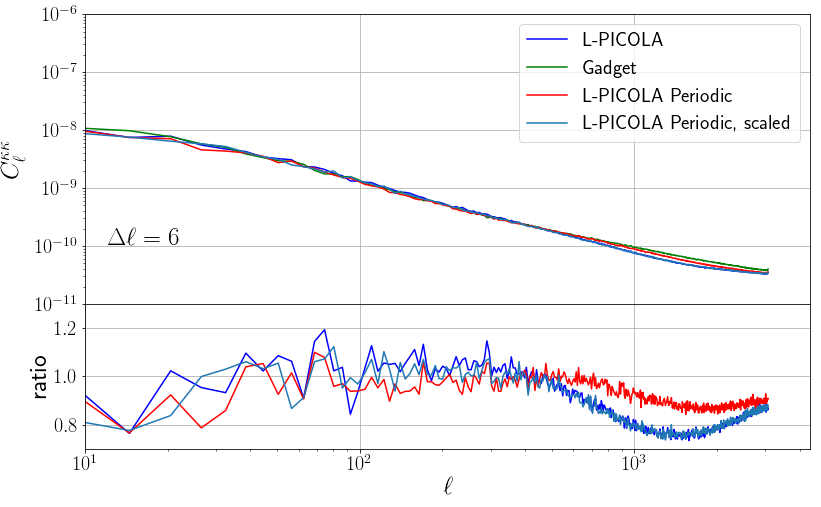}
\caption{Comparision of the power spectra of a Gadget 2 simulation and L-PICOLA simulations with different settings. All spectra were binned with bins of width $\Delta \ell = 6$ to reduce the noise. \label{f:powerspectra}}
\end{figure}

\begin{figure}[]
\centering
\includegraphics[width=0.75\textwidth]{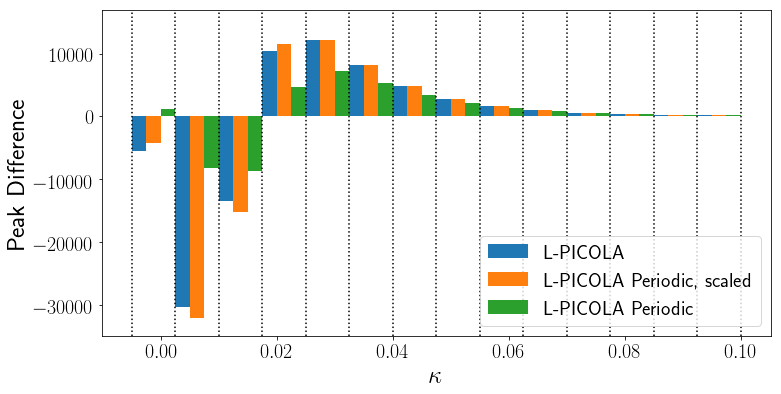}
\caption{Comparision of the peak count difference of convergence maps produced out of L-PICOLA simulations with different settings and a Gadget-2 simulation. \label{f:kappadist}}
\end{figure}

One disadvantage of the L-PICOLA code is a lower accuracy in small scale clustering \cite{Howlett2015, Sgier2018}. One way to improve this accuracy is to increase the number of mesh points of the L-PICOLA simulations. This, however, leads to a much higher memory usage and computation time. In figures \ref{f:powerspectra} and \ref{f:kappadist} we compare the power spectra and peak count difference of a Gadget-2 simulation and L-PICOLA simulations with and without the use of periodic boxes from our fiducial cosmology. The peak count difference was calculated using the unsmoothed full sky convergence maps. All maps were generated with the source redshift distribution given by equation \ref{eq:nz1}. The settings of the Gadget 2 simulation are explained in section \ref{s:nobody}. To measure the influence of periodic boxes we generated one L-PICOLA simulation using $1024^3$ particles and $2048^3$ mesh points without using periodic boxes. The second L-PICOLA simulation was done using correctly scaled periodic boxes, meaning we used 64 periodic boxes using $256^3$ particles and $512^3$ mesh points. Building the bigger boxes out of these smaller boxes one therefore recovers the same number of particles and mesh points as in the previous simulation. The third L-PICOLA simulation was done using the settings we decided to use eventually with $256^3$ particles and $1024^3$ mesh points. For the last L-PICOLA simulation we increased the number of mesh points to increase the small scale clustering accuracy.
Figures \ref{f:powerspectra} and \ref{f:kappadist} clearly show that the Gadget-2 code leads to much denser regions than all three L-PICOLA simulations. However one can also see that the use of correctly scaled periodic boxes does not have a significant influence on the peak count difference. Using periodic boxes with a higher number of mesh points leads to a smaller difference in both the power spectrum and the peak count difference. We therefore decided to use this setting for our simulations.

\section{Convergence Map Generation}
\label{s:convergence}

Here we give an quick overview of our procedure to generate convergence maps using \texttt{Ufalcon} \cite{Sgier2018}. A more detailed approach can be found in \cite{Sgier2018}. Using the connection between the convergence and the overdensity, the convergence at a given pixel $\theta_\mathrm{pix}$ can be calculated using
\begin{equation}
\kappa(\theta_\mathrm{pix}) \approx \frac{3}{2}\Omega_\mathrm{m}\sum_bW_b\frac{H_0}{c}\int_{\Delta z_b}\frac{c\mathrm{d}z}{H_0E(z)}\delta\left(\frac{c}{H_0}\mathcal{D}(z)\hat{n}_\mathrm{pix},z\right),
\end{equation}
where $\mathcal{D}(z)$ is the dimensionless comoving distance, $\hat{n}_\mathrm{pix}$ is a unit vector pointing to the pixels center and $E(z)$ is given by
\begin{equation}
\mathrm{d}\mathcal{D} = \frac{\mathrm{d}z}{E(z)}.
\end{equation}
The sum runs over all redshift shell and $\Delta z_b = 0.01$ is the thickness of the shell. Each shell gets the additional weight $W_b$ which depends on the redshift distribution of the source galaxies. For a Dirac shaped source distribution at $z_s$ the weight can be calculated using
\begin{equation}
W^{\mathrm{delta}}_b = \left(\int_{\Delta z_b}\frac{\mathrm{d}z}{E(z)}\frac{\mathcal{D}(z)\mathcal{D}(z,z_s)}{\mathcal{D}(z_s)}\frac{1}{a(z)}\right)/\left(\int_{\Delta z_b}\frac{\mathrm{d}z}{E(z)}\right).
\end{equation}
This can be generalized for any given redshift distribution $n(z)$ of source galaxies
\begin{equation}
W^{n(z)}_b = \left(\int_{\Delta z_b}\frac{\mathrm{d}z}{E(z)}\int_z^{z_s}\mathrm{d}z'n(z')\frac{\mathcal{D}(z)\mathcal{D}(z,z')}{\mathcal{D}(z')}\frac{1}{a(z)}\right)/\left(\int_{\Delta z_b}\frac{\mathrm{d}z}{E(z)}\int_{z_0}^{z_s}\mathrm{d}z'n(z')\right),
\end{equation}
where $z_0$ is the redshift of the first shell and $z_s$ the redshift of the last shell that is added. In this work we always used the redshift boundaries $z_0 = 0.1$ and $z_s = 1.5$.

\section{Choice of the Mock Observation}
\label{s:cherrypicking}

In a real experiment one usually has only one data set. In this theoretical work we could however choose our mock observation out of 2600 (3900 for the tomographic run) noisy realizations from our fiducial cosmology. To make a justified choice of our mock observation we decided to compute the figures-of-merit $\mathrm{FoM}_{\sigma_8-\Omega_m}$ for all 2600 noisy realizations from our non tomographic and  for all 3900 from our tomographic run. We found that the average FoM for the non tomographic run was $\left\langle \mathrm{FoM}_{\sigma_8-\Omega_m} \right\rangle = 947 \pm 137$ and $\left\langle \mathrm{FoM}_{\sigma_8-\Omega_m} \right\rangle = 887 \pm 129$ for the tomographic run. We decided to choose a mock measurement that had a FoM close to the mean of the distribution.


\end{document}